\documentclass{iau}

\usepackage{graphicx}

\begin{document}

\title[Einstein-Gau\ss -Bonnet-dilaton black holes] 
{Black holes\\ in Einstein-Gau\ss -Bonnet-dilaton theory}

\author[Bl\'azquez-Salcedo et al.]   
{Jose Luis Bl\'azquez-Salcedo $^1$, 
Vitor Cardoso $^2$,
Valeria Ferrari $^3$, \\
Leonardo Gualtieri $^3$,
Panagiota Kanti$^4$,
Fech Scen Khoo $^5$, \\
Burkhard Kleihaus $^1$,
Jutta Kunz $^1$, 
Caio F. B. Macedo$^6$, \\
Sindy Mojica $^7$,
Paolo Pani $^3$
\and Eugen Radu $^8$}

\affiliation{$^1$Institut f\"ur  Physik, Universit\"at Oldenburg,
Postfach 2503, D-26111 Oldenburg, Germany \\[\affilskip]
$^2$CENTRA, Departamento de F\'{\i}sica, Instituto Superior T\'ecnico, 
Universidade de Lisboa\\
Avenida~Rovisco Pais 1, 1049 Lisboa, Portugal \\[\affilskip]
$^3$Dipartimento di Fisica, ``Sapienza'' Universit\`a di Roma \& Sezione 
INFN Roma1\\
Piazzale Aldo Moro 5, 00185 Roma, Italy \\[\affilskip]
 $^4$ Division of Theoretical Physics, Department of Physics, University of Ioannina \\ Ioannina GR-45110, Greece \\[\affilskip]
$^5$Department of Physics and Earth Sciences, Jacobs University,
D-28759 Bremen, Germany \\[\affilskip]
$^6$Faculdade de F\'isica, Universidade Federal do Par\'a  66075-110 Bel\'em, Par\'a, Brazil \\[\affilskip]
$^7$Escuela de F\'isica, Universidad Industrial de Santander \\
A. A. 678, Bucaramanga 680002, Colombia \\[\affilskip]
$^8$Departamento de F\'isica da Universidade de Aveiro and CIDMA\\
 Campus de Santiago, 3810-183 Aveiro, Portugal 
 }

\pubyear{2016}
\volume{324}  
\setcounter{page}{1}
\jname{New Frontiers in Black Hole Astrophysics}
\editors{A.C. Editor, B.D. Editor \& C.E. Editor, eds.}


\maketitle

\begin{abstract}
Generalizations of the Schwarzschild and Kerr black holes are 
discussed in an astrophysically viable generalized theory of gravity, 
which includes higher curvature corrections 
in the form of the Gauss-Bonnet term, coupled to a dilaton. 
The angular momentum of these black holes can slightly exceed the Kerr 
bound. The location and the orbital frequency of particles in their innermost stable circular 
orbits can deviate significantly from the respective Kerr values. 
Study of the quasinormal modes of the static black holes gives
strong evidence that they are mode stable against polar and axial
perturbations. Future gravitational wave observations should
improve the current bound on the Gauss-Bonnet coupling constant,
based on observations of the low-mass x-ray binary A 0620-00.
\end{abstract}

\firstsection 
\section{Introduction}

So far General Relativity (GR) has passed all tests 
in the Solar System and beyond.
Still we expect that GR will be superseded by a generalized theory of gravity,
which encompasses GR as a limit. Reasons for this expectation reside on the
one hand on the theoretical side, where the incompatibility of GR with quantum
mechanics has sparked the search for a theory of quantum gravity. On the other
hand, when GR is applied to the evolution of the Universe as a whole, the 
necessity for the presence of dark matter and dark energy arises, 
the nature of which is currently unknown.

By now a large number of alternative theories of gravity have been proposed
and some of their implications for astrophysical objects have been studied. 
The review {\sl Testing General Relativity with 
Present and Future Astrophysical Observations}
by \cite{Berti:2015itd}
gives a broad account of recent activities in this area.

Here we consider Einstein--Gauss-Bonnet-dilaton (EGBd) gravity,
which represents a very interesting and well motivated 
extension of GR.
The EGBd action is obtained by adding 
a real scalar field, a dilaton, to the GR action
which is non-minimally coupled to the Gauss-Bonnet (GB) term.
The resulting theory of gravity has quadratic curvature
terms, but it has only second order equations of motion.
Moreover, the EGBd action arises naturally in the framework
of the low-energy effective string theories
(see $e.g.$ \cite{Moura:2006pz}).

To represent a viable alternative theory of gravity,
EGBd theory must satisfy theoretical and observational constraints.
These yield bounds on the GB coupling constant $\alpha$.
Solar system measurements, as e.g., measurements of the Shapiro time delay,
yield the rather weak bound,
$\sqrt{\alpha} \lesssim 10^{13} {\rm cm}$
(\cite{Bertotti:2003rm}).
In contrast, black holes in low-mass x-ray binaries
can give a much stronger bound,
since black holes in EGBd theory carry a scalar charge.
Observations of the black hole low-mass x-ray binary A0620-00
(\cite{Cantrell:2010vh})
then lead to a constraint 
based on the orbital decay rate
(\cite{Yagi:2012gp})
\begin{equation}
\sqrt{\alpha} \lesssim 10^{6} {\rm cm} \ .
\label{boundY}
\end{equation}
From the theoretical side the 
mere existence of black hole solutions implies
an upper bound (\cite{Kanti:1995vq,Pani:2009wy})
\begin{equation}
\frac{\alpha}{M^2} \lesssim 0.691 \ .
\label{boundK}
\end{equation}
Future observations of quasi-periodic oscillations (QPOs) from accreting black holes could be used to further constrain the theory (\cite{Maselli:2015}).

Here we explore the physical properties of static 
and rotating EGDd black holes 
(\cite{Mignemi:1992nt,Kanti:1995vq,Torii:1996yi,Pani:2009wy,Kleihaus:2011tg,Pani:2011gy,Ayzenberg:2014aka,Kleihaus:2014lba,Maselli:2015tta,Kleihaus:2015aje,Blazquez-Salcedo:2016enn}).
In particular, we study their domain of existence and point out
their differences to the Schwarzschild and Kerr black holes of GR.
By computing the quasinormal modes of the static EGBd black holes,
we infer their mode stability. Finally we address how future
gravitational wave observations may improve the current bound on the
GB coupling constant.

\section{Einstein-Gauss-Bonnet-dilaton gravity}

The action of EGBd gravity is given by
\begin{equation}
S=\frac{1}{16 \pi}\int d^4x \sqrt{-g} \left[R - \frac{1}{2}
 (\partial_\mu \phi)^2
        + \frac{\alpha}{4} e^{\phi} R^2_{\rm GB}   \right] \ , 
\end{equation}
where $R$ is the curvature scalar, $\phi$ is a real scalar field, 
and $\alpha$ is the GB coupling constant.
The dilaton coupling constant has been given its
string theory value throughout,
and $R^2_{\rm GB}$ is the Gauss-Bonnet term,
which is quadratic in curvature
\begin{equation}
R^2_{\rm GB} = R_{\mu\nu\rho\sigma} R^{\mu\nu\rho\sigma}
                    - 4 R_{\mu\nu} R^{\mu\nu} + R^2 \ .
\end{equation}

The resulting set of equations of motion are of second order.
The ``modified'' Einstein equations read
\begin{equation}
G_{\mu\nu}  = \frac{1}{2}\ T_{\mu\nu}
\end{equation}
with the effective stress-energy tensor
\begin{eqnarray}
T_{\mu\nu} & = & T_{\mu\nu}^{(\phi)}+
  \frac{\alpha}{2} e^{\phi} T_{\mu\nu}^{\rm (GBd)} \ , 
\end{eqnarray}
{where}
\begin{eqnarray}
T_{\mu\nu}^{(\phi)} &=& \nabla_\mu \phi \nabla_\nu \phi
                 -\frac{1}{2}g_{\mu\nu}\nabla_\lambda \phi \nabla^\lambda\phi
\\
T_{\mu\nu}^{\rm (GBd)} &=&
H_{\mu\nu}
  +4\left(\nabla^\rho \phi \nabla^\sigma \phi
           + \nabla^\rho\nabla^\sigma \phi\right) P_{\mu\rho\nu\sigma}
\ , 
\end{eqnarray}
where $H_{\mu\nu}$ is quadratic and $P_{\mu\rho\nu\sigma}$ is linear in the 
curvature.

The dilaton equation is given by
\begin{equation}
\nabla^2 \phi  =  \frac{\alpha}{4}
e^{\phi}R^2_{\rm GB} \ .
\end{equation}
Because of the contributions from the GB term on the right hand side
of the Einstein equations,
the theory allows for negative ``effective'' energy densities
giving rise to black holes with scalar ``hair'', although secondary. 
Moreover, this theory allows for wormholes without 
the need for exotic matter
(\cite{Kanti:2011jz}).

\section{Black hole properties}

The vacuum black holes in GR consist of the family of 
static Schwarzschild and rotating Kerr black holes, 
which in general form the basis of current analyses of
astrophysical observations.
The GR black holes possess very special properties,
in particular, all Kerr black holes are uniquely
described by only two parameters, their mass $M$ and 
their angular momentum $J$. 
For Kerr black holes
the ratio $j=J/M^2$ is bounded, $|j|\le 1$, 
with the extremal black holes saturating the limit.

{\underline{\it Static EGBd black holes}}.

The full set of static EGBd black holes was first constructed by 
\cite{Kanti:1995vq}.
In this analysis it was realized, that unlike GR black holes
the EGBd black holes possess a lower bound for the size
and for the mass, when the GB coupling is kept fixed,
i.e., for a given theory.
This bound arises when the metric and dilaton functions
are expanded at the horizon $r_h$, since the expansion for the dilaton
involves a coefficient with a square root
\begin{equation}
\sqrt{1-6 \frac{\alpha^2}{r_h^4} e^{2 \phi_h}} \ , 
\end{equation}
whose radicand should not be negative in order 
to retain a real value for the scalar field.

\begin{figure}[t]
\begin{center}
\mbox{
 \includegraphics[width=1.8in,angle=270]{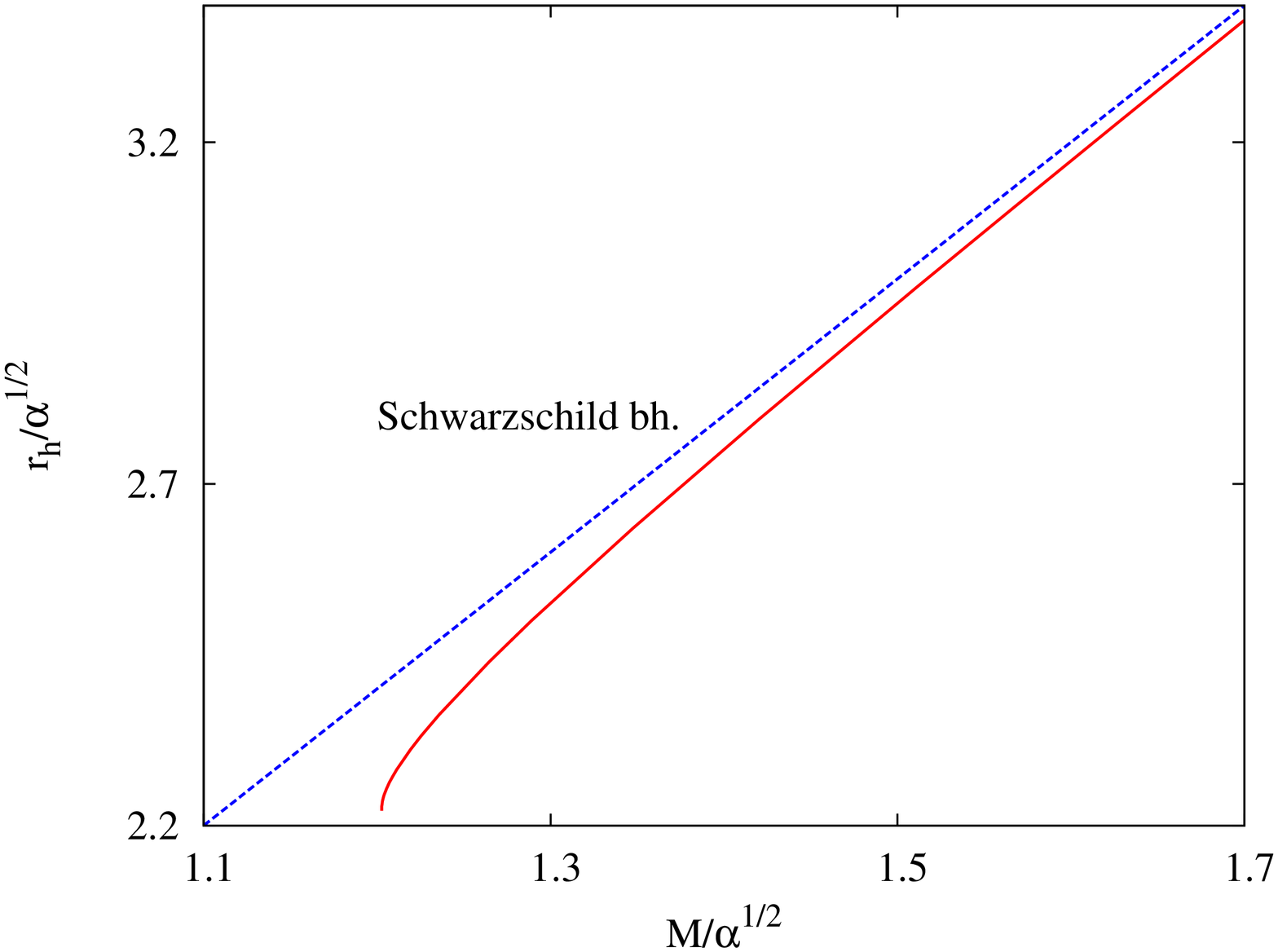}
 \includegraphics[width=1.8in,angle=270]{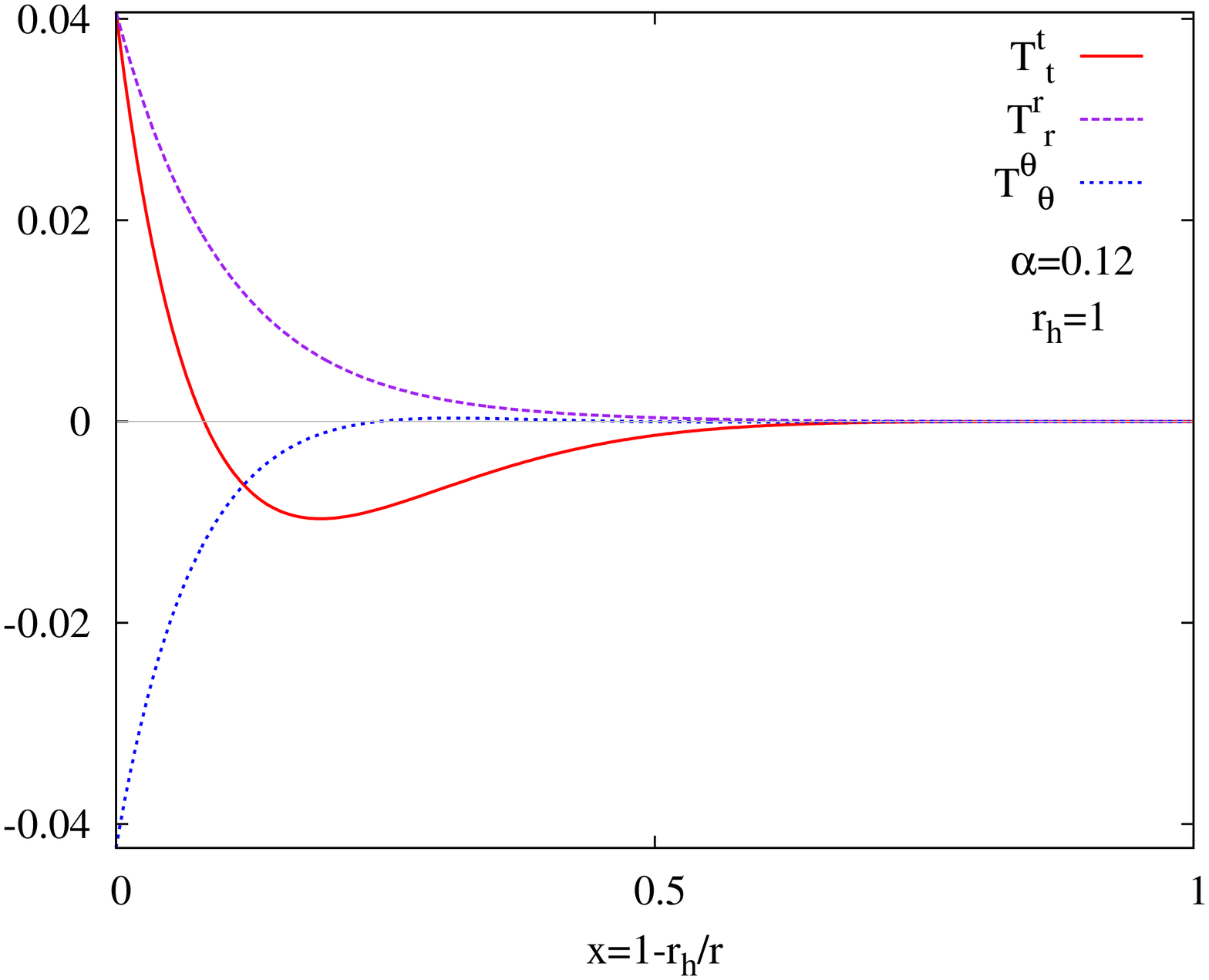}
}
 \caption{Scaled horizon radius $r_h$ versus
scaled mass $M$ of static EGBd black holes
with comparison to Schwarzschild black holes (left).
Profile of the components of the effective stress-energy tensor, $T_t^t$, $T_r^r$ and $T_\theta^\theta$,
for a typical static EGBd black hole (right).}
   \label{fig1}
\end{center}
\end{figure}

We demonstrate the dependence of the horizon radius $r_h$
on the mass $M$ for the static EGBd black holes in
Fig.\,\ref{fig1} 
in the vicinity of the minimal mass and the minimal radius.
It is the presence of the minimal mass,
which leads to the bound (\ref{boundK}).
We also note from the figure, that whereas for a given $\alpha$
the deviations from Schwarzschild are quite pronounced
for small masses, i.e., the EGBd black holes have a much smaller
area for the same mass, the EGBd solutions tend toward
the Schwarzschild solutions for larger masses.

Fig.\,\ref{fig1} 
also exhibits the relevant components of the
stress-energy tensor of a typical static black hole. 
It reveals that the energy density $\rho=-T^t_t$ is
negative in the vicinity of the horizon.
This allows these static black holes to
circumvent the no-hair theorem and carry scalar hair.

{\underline{\it Rotating EGBd black holes}}.

\begin{figure}[t]
\begin{center}
 \includegraphics[width=2.6in]{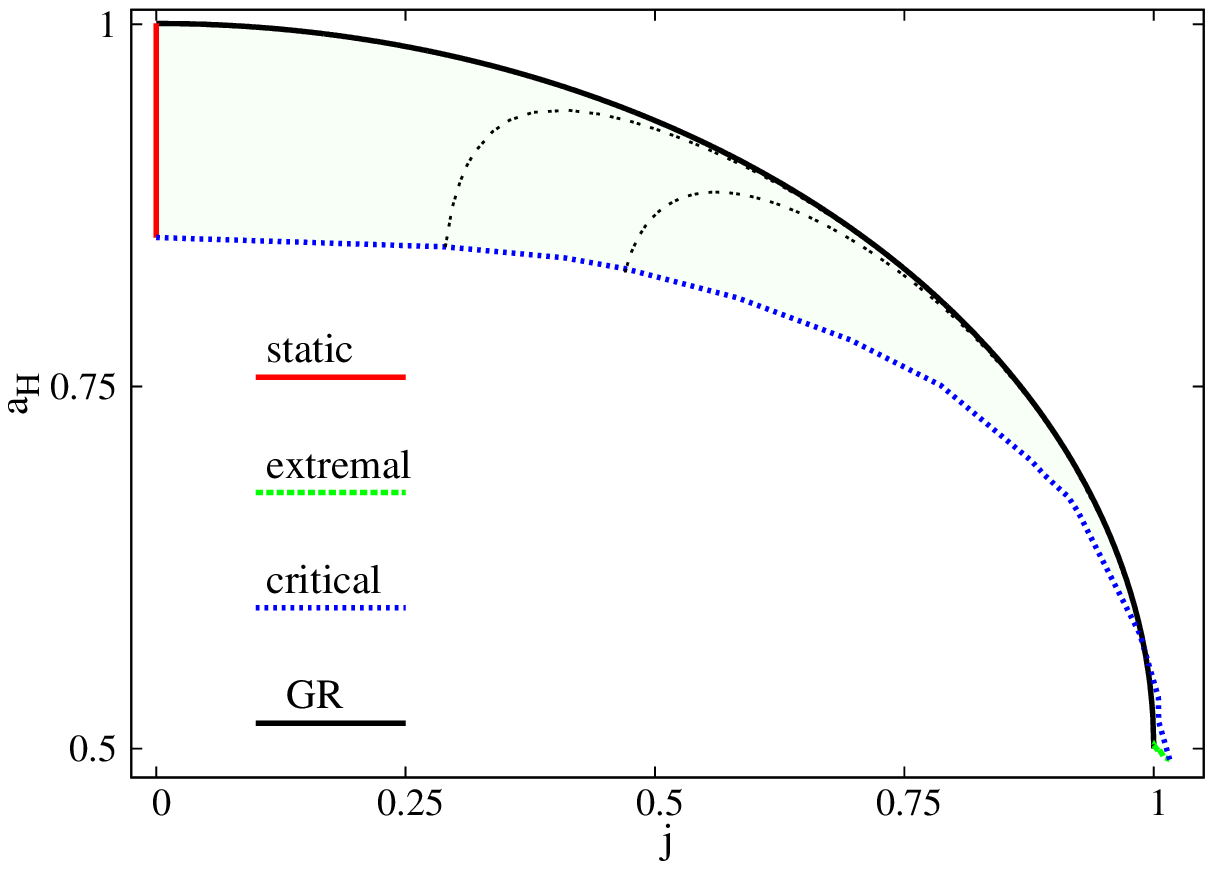}
 \includegraphics[width=2.6in]{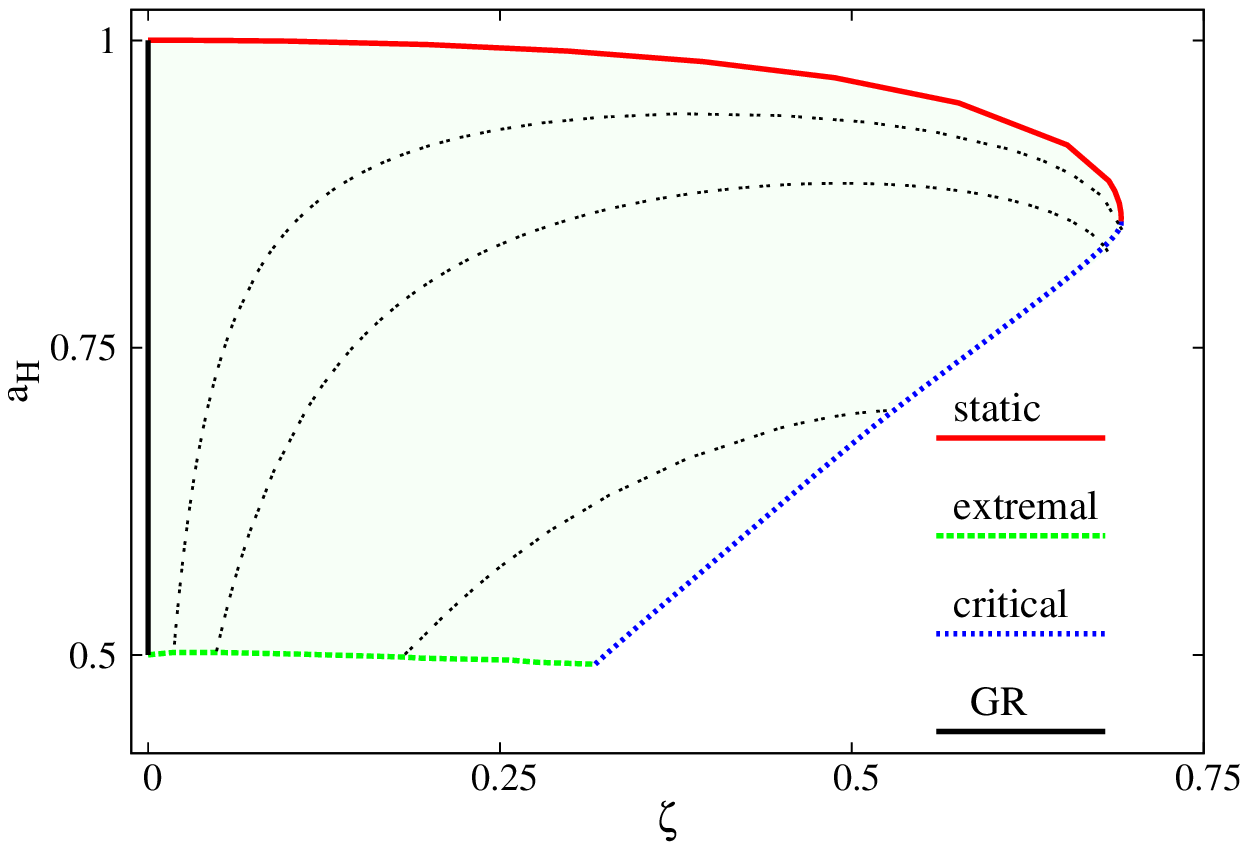}
 \caption{The scaled area versus the scaled angular momentum
(left). The scaled area versus the scaled GB coupling constant (right).
Also indicated are the static, the extremal, the critical
and the GR black holes.}
   \label{fig2}
\end{center}
\end{figure}

The inclusion of rotation is essential for astrophysical applications,
but also from a theoretical point of view. The domain of
existence of rotating EGBd black holes has been explored 
in \cite{Kleihaus:2011tg,Kleihaus:2015aje} 
and is exhibited in Fig.\,\ref{fig2}.
This domain of existence is limited by the static EGBd black holes,
the Kerr black holes, the extremal rotating EGBd black holes
and the set of critical EGBd black holes.
The latter arise analogous to the static case, when the radicand of
a square root in the horizon expansion vanishes.

On the left of Fig.\,\ref{fig2} the scaled area is exhibited
versus the scaled angular momentum. It comes as a surprise,
that EGBd theory allows for black holes violating the Kerr bound,
i.e., possessing $|j| > 1$, although this violation is small.
On the right of Fig.\,\ref{fig2} for comparison the scaled
area is shown versus the scaled GB parameter $\zeta=\alpha/M^2$.

\begin{figure}[t]
\begin{center}
 \includegraphics[width=2.6in]{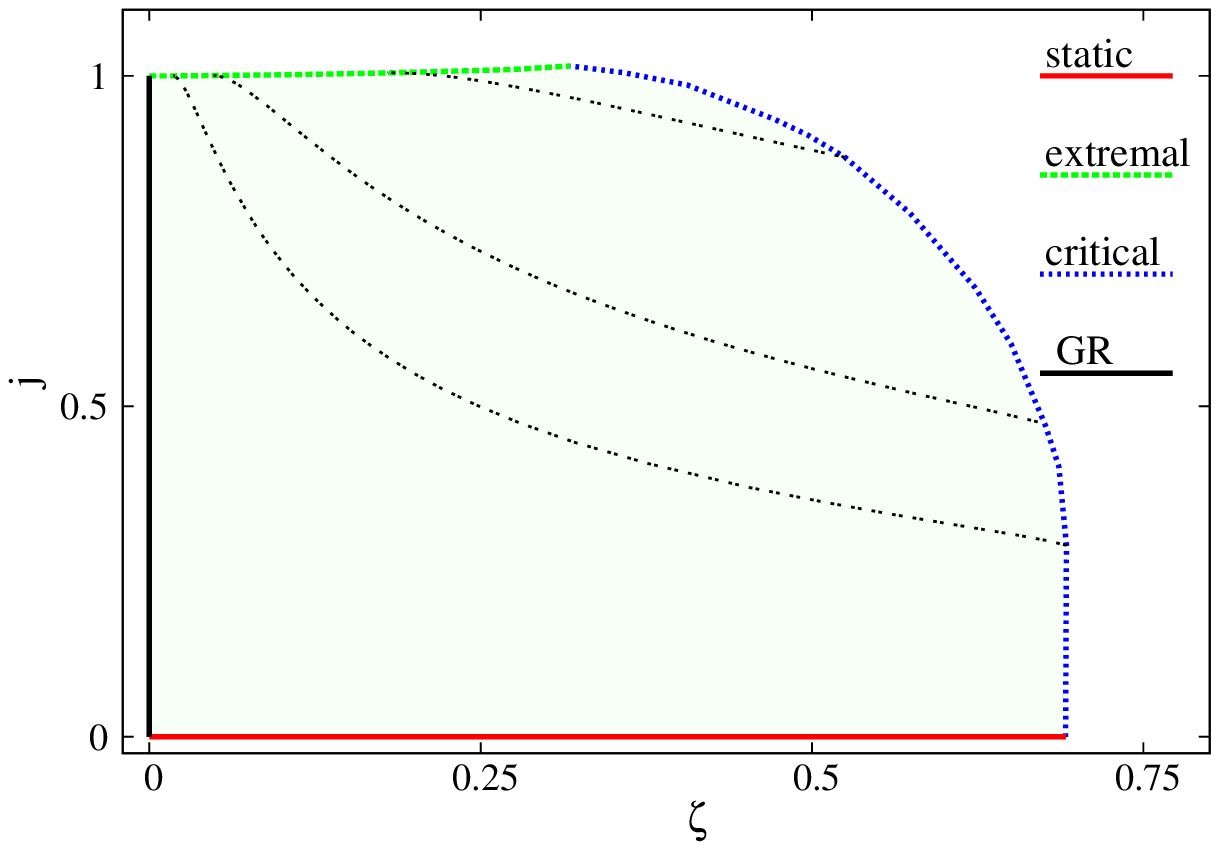}
 \includegraphics[width=2.6in]{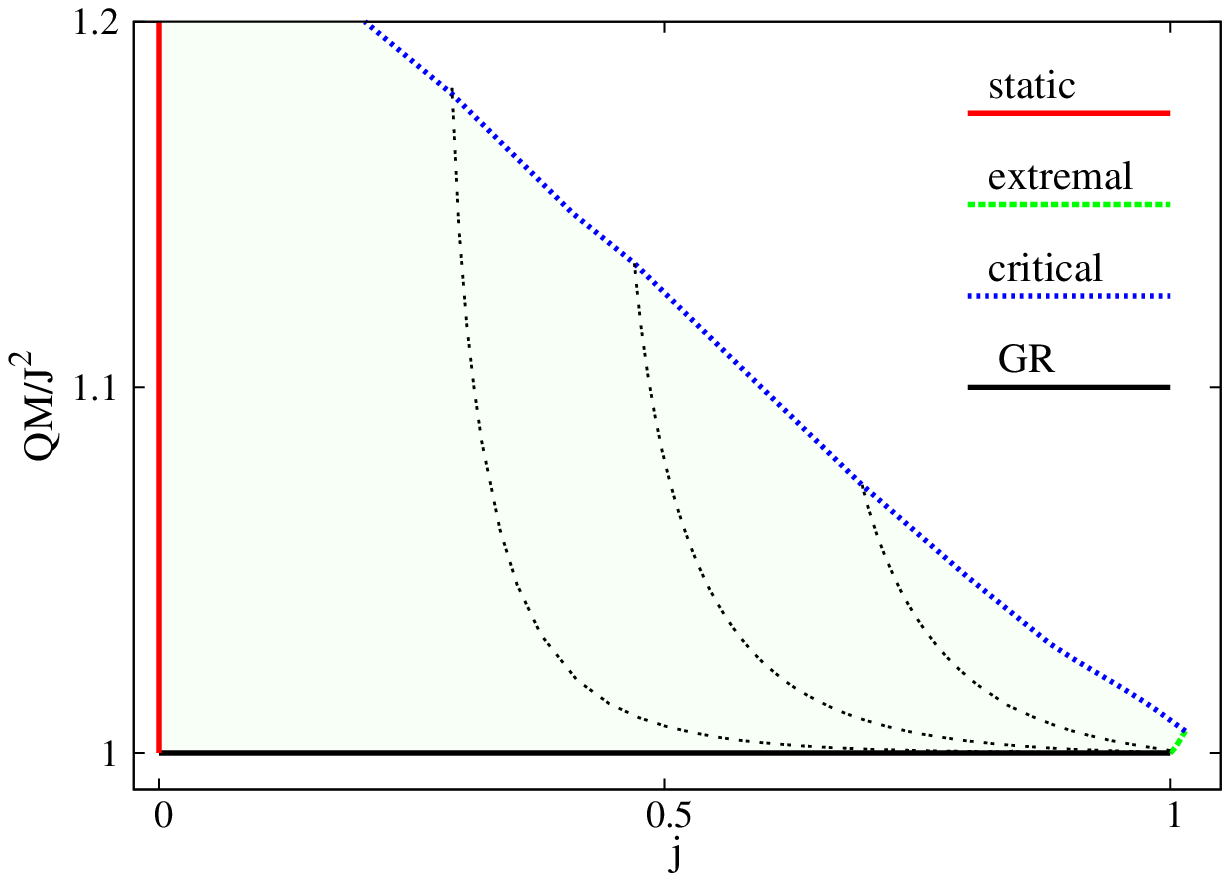}
 \caption{The scaled angular momentum 
versus the scaled GB coupling constant (left).
The scaled quadrupole moment
versus the scaled angular momentum (right).
Also indicated are the static, the extremal, the critical
and the GR black holes.}
   \label{fig3}
\end{center}
\end{figure}

As seen on the left of Fig.\,\ref{fig3},
the scaled angular momentum reaches its maximum roughly at half
the maximal value of the scaled GB coupling, where the critical and the
extremal rotating EGBd black holes merge. We note that the extremal rotating EGBd black holes 
possess a regular metric on the horizon, whereas the dilaton field
diverges on the horizon at the poles.

The quadrupole moment of the Kerr black holes is completely fixed by
the global charges, $Q=-J^2/M$.
Fig.\,\ref{fig3} (right), however, shows, that the quadrupole moment of
EGBd black holes can considerably differ from the Kerr case.
The Kerr values for the scaled moment of inertia are given by
$J/(\Omega_{\rm H} M^3) = 2(1+\sqrt{1-j^2})$,
thus they are fixed by the value of $j$.
For the EGBd black holes this is no longer the case, where they 
decrease monotonically from the Kerr value for a fixed $j$.

{\underline{\it Geodesics}}.

\begin{figure}[t]
\begin{center}
 \includegraphics[width=2.6in]{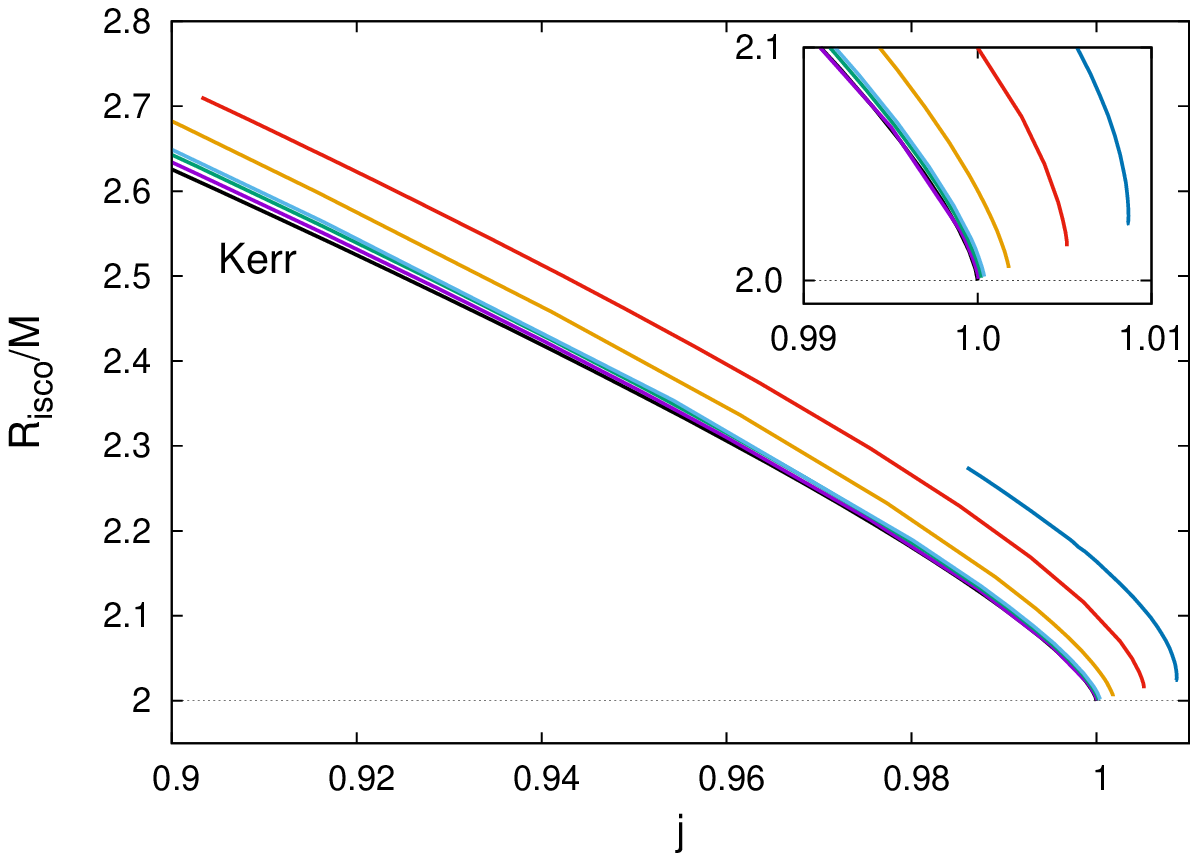}
 \includegraphics[width=2.6in]{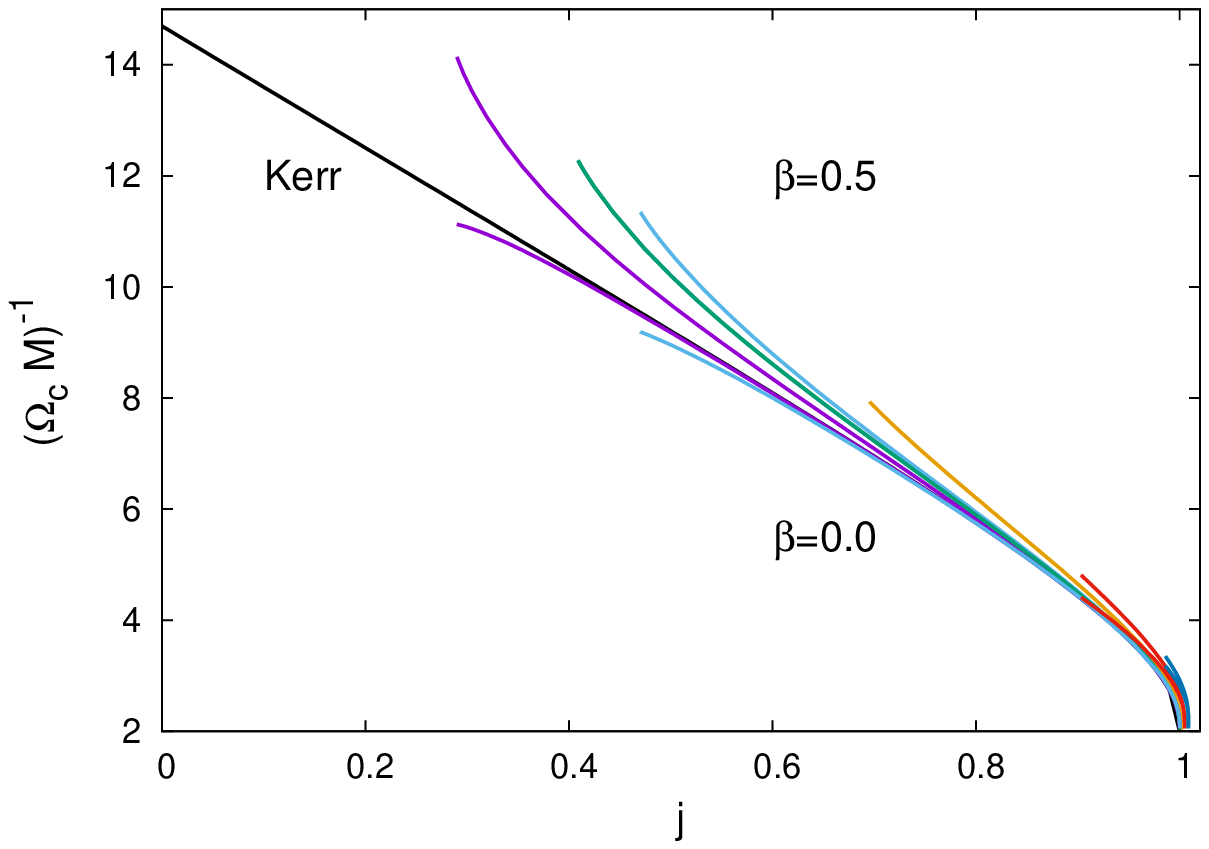}
 \caption{The scaled ISCO radius versus the scaled angular momentum (left).
The inverse of the scaled ISCO frequency 
versus the scaled angular momentum for matter coupling constant
$\beta=0$ and $\beta=1/2$ (right).
The curves correspond to different horizon velocities of the black hole.}
   \label{fig4}
\end{center}
\end{figure}

Also the study of the geodesics of these black holes is both
of intrinsic interest and of astrophysical relevance.
The analysis of the geodesics is based on the Lagrangian
\begin{equation}
{\cal L} = \frac{1}{2} e^{-2{\beta} \phi} g_{\mu\nu} 
\dot{x}^\mu \dot{x}^\nu \ ,
\end{equation}
where $\beta$ is a coupling constant  ($\beta=1/2$ for string theory).
Since the general geodesic equations are not separable (see the discussion
in \cite{Kleihaus:2015aje}, showing the EGBd black holes are of
Petrov type I), it is simpler to consider equatorial
motion. In particular, we now restrict the analysis
to the innermost stable circular orbits (ISCOs) of these black holes.

A perturbative analysis for slow rotation was performed in 
\cite{Pani:2009wy}, while the analysis of the general rotating case
was performed in \cite{Kleihaus:2011tg,Kleihaus:2015aje}.
We exhibit the scaled ISCO radius versus the scaled angular momentum 
in Fig.\,\ref{fig4} (left) for coupling constant $\beta=1/2$.
(The ISCO radius here represents the circumferential radius,
not the Boyer-Lindquist coordinate radius.)
We note from the figure that the ISCO radius of EGBd black holes
exceeds the one of Kerr black holes.
Also shown in the figure is the inverse of the scaled ISCO frequency 
versus the scaled angular momentum (right).
As compared to the Kerr case, the frequencies are smaller
when $\beta=1/2$. For $\beta=0$, however, they are larger.

\

{\underline{\it Quasinormal modes and gravitational waves}}.

\begin{figure}[t]
\begin{center}
\mbox{
 \includegraphics[width=1.6in,angle=270]{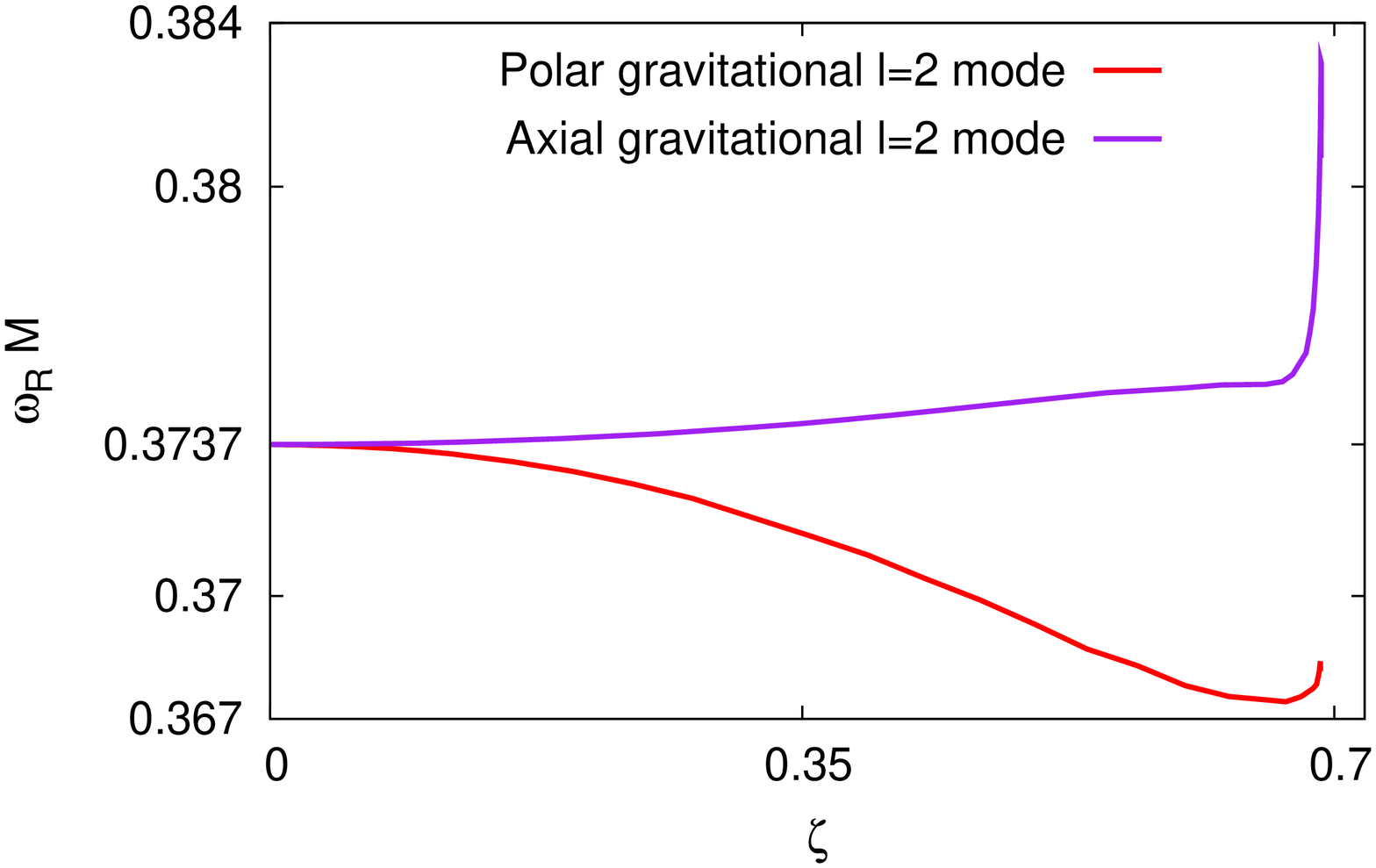}
 \includegraphics[width=1.6in,angle=270]{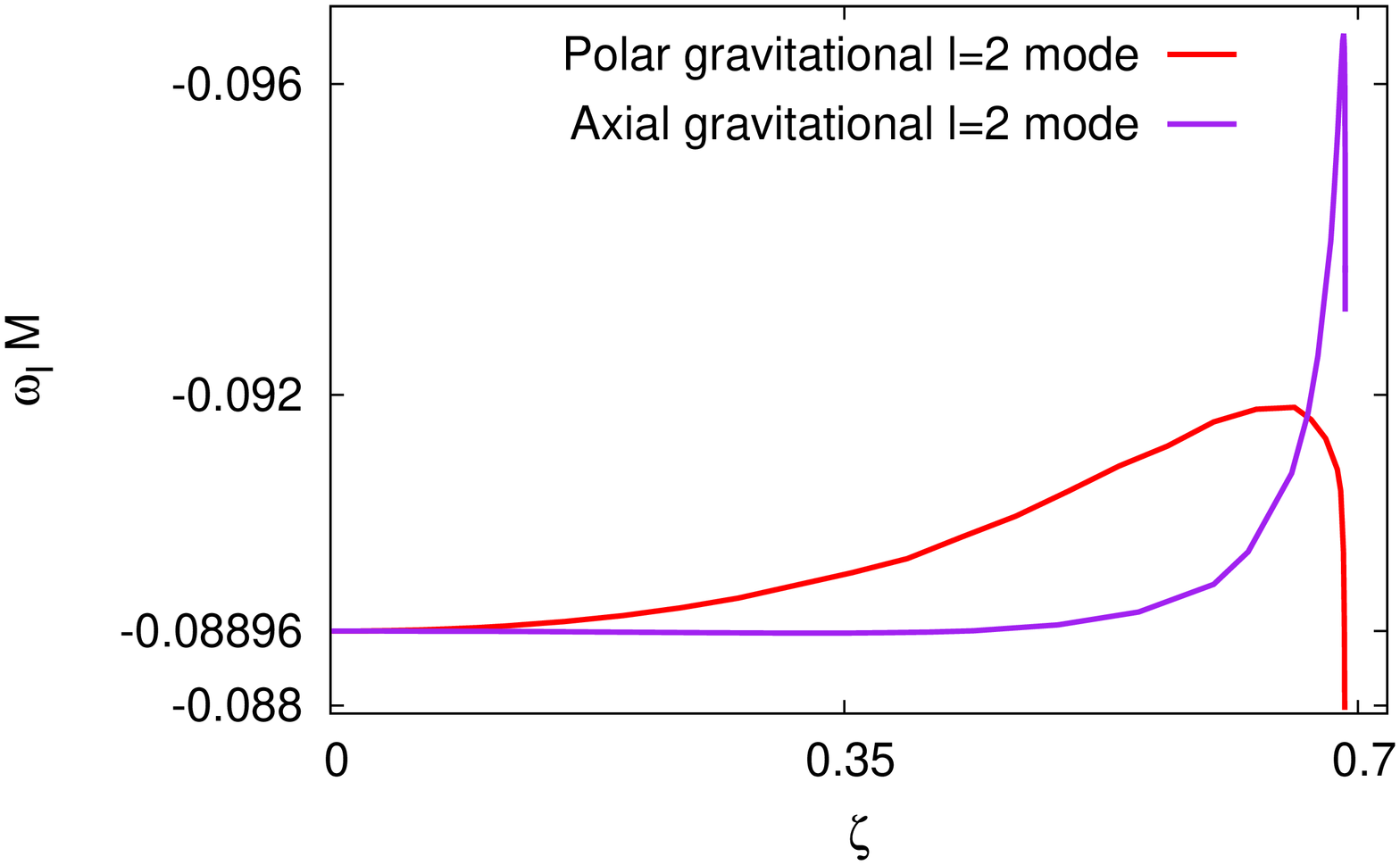}
}
 \caption{Real (left) and imaginary (right) parts of the 
gravita\-tional axial and gravitational-led polar $l=2$ 
fundamental modes. 
The Schwarzschild values are recovered at $\zeta=0$.}
   \label{fig5}
\end{center}
\end{figure}

\begin{figure}[t]
\begin{center}
\mbox{
 \includegraphics[width=1.6in,angle=270]{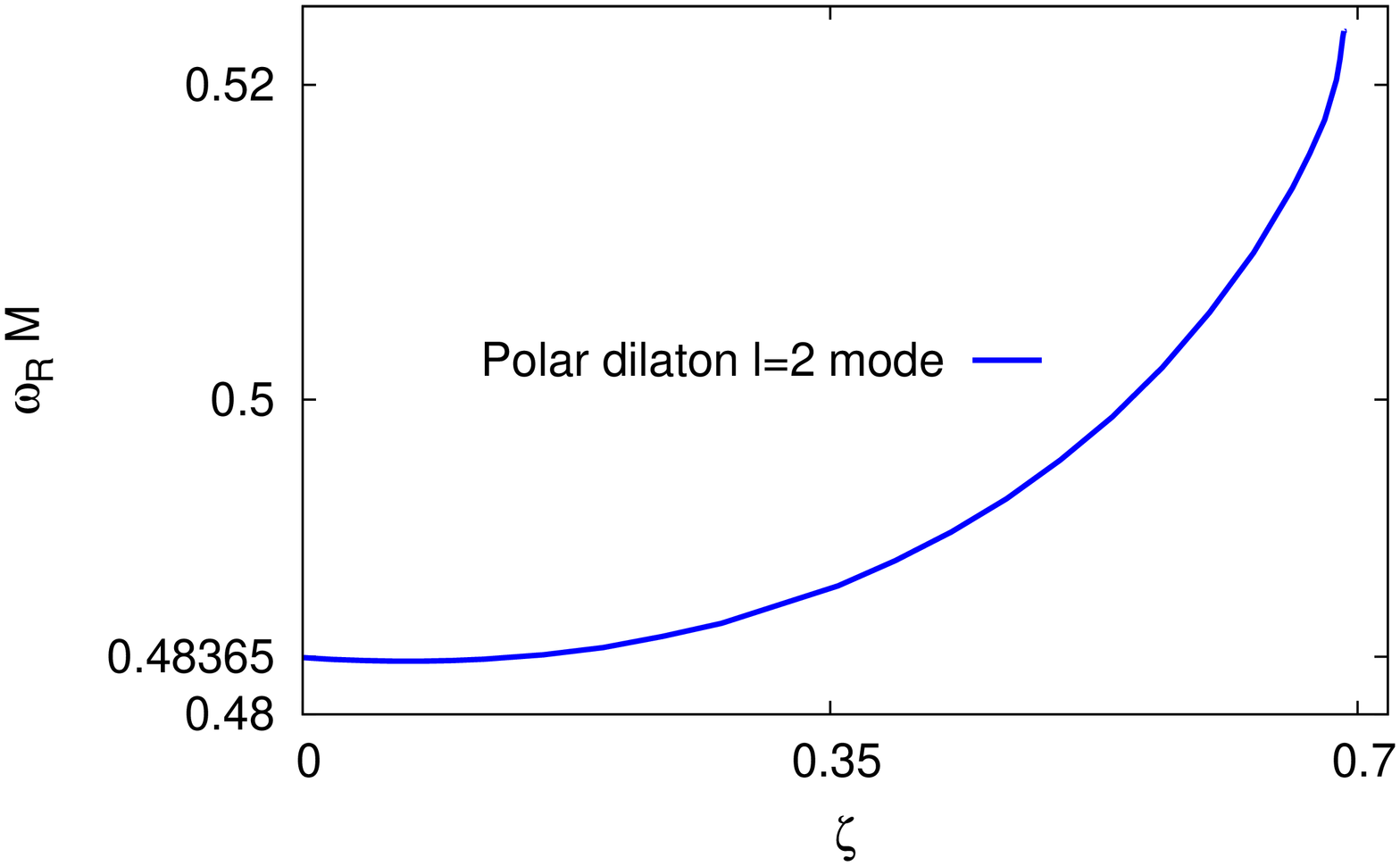}
 \includegraphics[width=1.6in,angle=270]{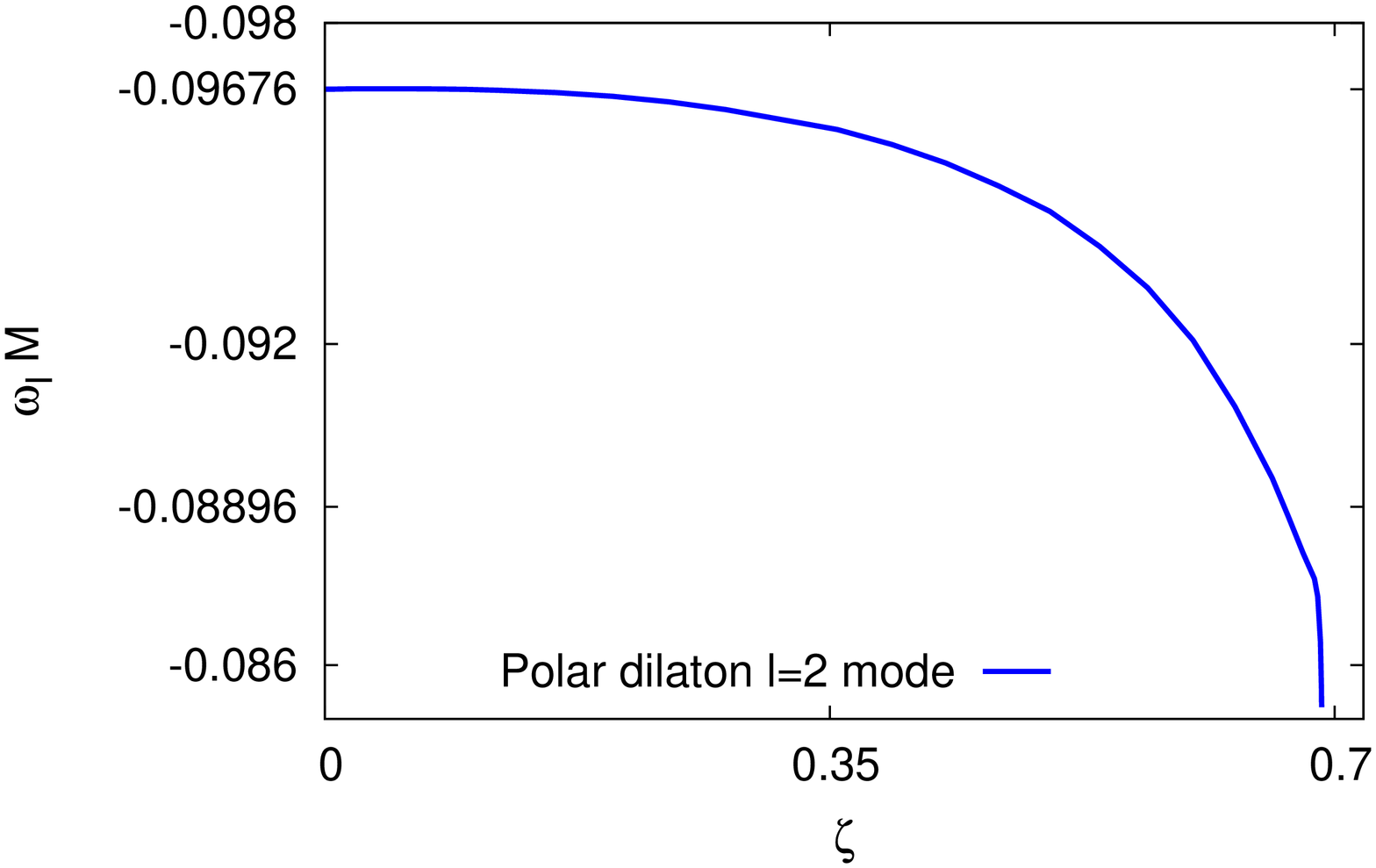}
}
 \caption{Real (left) and imaginary (right) parts of the 
scalar-led polar $l=2$ fundamental modes.
The Schwarzschild values are recovered at $\zeta=0$.}
   \label{fig6}
\end{center}
\end{figure}

When a black hole is perturbed it reacts with the emission of
gravitational waves, to settle down again in a stationary state.
In an astrophysical setting such perturbations can for instance
arise through the infall of a smaller mass, or through
the coalescence of black holes. 
Observation of the latter process was announced
earlier this year by the LIGO collaboration
\cite{Abbott:2016blz}. The process consists of three major phases,
the inspiral, the merger and the ringdown.
We will now briefly address 
the quasinormal modes of EGBd black holes,
relevant for the ringdown phase.

In the following we restrict to the quasinormal modes of static
EGBd black holes.
The perturbations of the metric and the scalar field
are given by
\begin{eqnarray}
g_{ab}&=g_{ab}^{(0)}+\varepsilon\, h_{ab} \, , \, \, \,
\phi&=\phi_0(r)+\varepsilon\,\delta\phi \ ,
\end{eqnarray}
where the subscript zero denotes the unperturbed fields.
They can be expanded in terms of spherical harmonics
and Fourier transformed, where $\omega$ is the Fourier frequency.

To determine the quasinormal modes, we first note, that
the modes decouple according to their
behavior under parity transformations.
The axial modes involve only the metric,
while the polar modes involve the dilaton field as well.
Imposing physical boundary conditions on the modes,
such that the solutions are purely ingoing at the horizon
and purely outgoing at infinity, the coupled
equations can be solved for the axial and the polar modes
(\cite{Blazquez-Salcedo:2016enn}).

We exhibit the frequencies of the lowest
fundamental axial modes in Fig.\,\ref{fig5},
where the real part is shown on the left and the imaginary part on
the right, versus the scaled GB coupling constant $\zeta$.
The Schwarzschild values are recovered at $\zeta=0$.
The deviations from the Schwarzschild values are small
for $\zeta \le 0.3$, but increase significantly
in the vicinity of the critical solution.

For the polar modes we have to distinguish between gravitational-led modes
and scalar-led modes.
In the limit $\zeta \to 0$
the gravitational-led EGBd modes
reduce to the gravitational modes of the Schwarzschild metric, 
while the scalar-led EGBd modes reduce to the modes
of a test scalar field in the Schwarzschild background.
Fig.\,\ref{fig5} also shows the frequencies of the lowest
gravitational-led EGBd modes, while the scalar-led modes
are shown in Fig.\,\ref{fig6}.
As seen in the figures,
the deviations from the Schwarzschild modes are somewhat
larger for the polar modes.

\section{Wormholes}

\begin{figure}[t]
\begin{center}
 \includegraphics[width=2.6in]{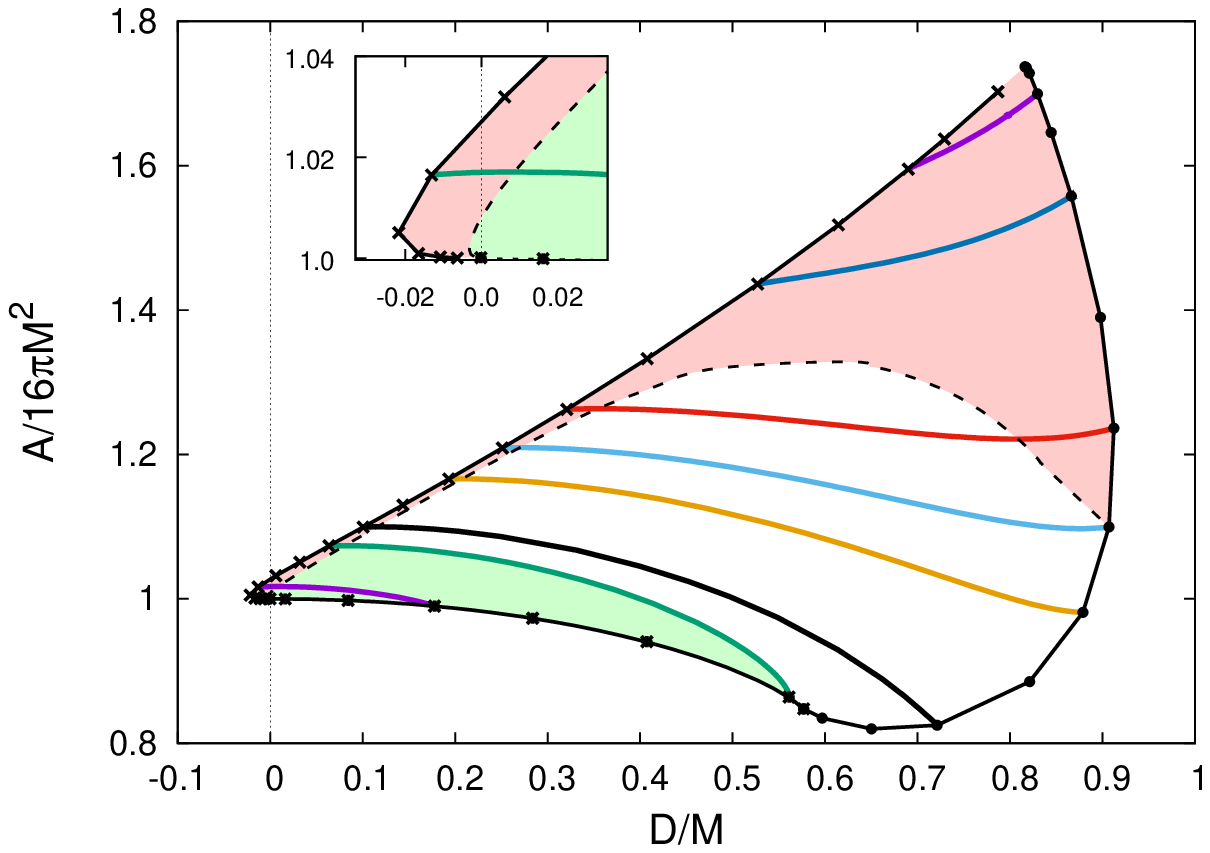}
 \includegraphics[width=2.6in]{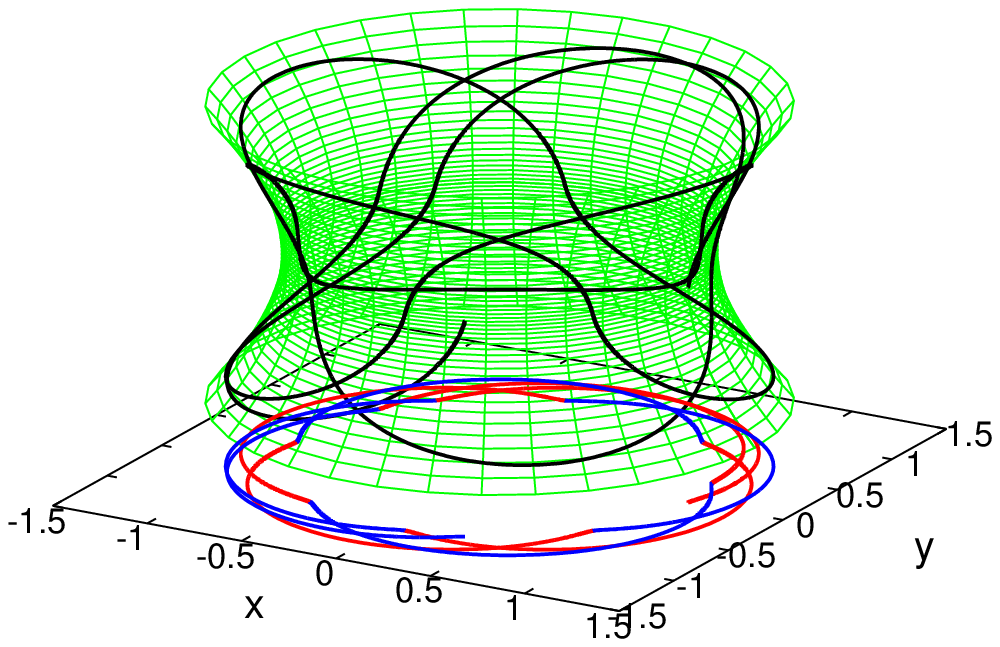}
 \caption{Domain of existence of wormholes in terms of the
scaled throat area versus the scaled scalar charge.
Wormholes in the lower part of the diagram are expected to be stable (left).
Bound orbit of a test particle traveling between the
universes (right).}
   \label{fig7}
\end{center}
\end{figure}

As mentioned in Sec. 2, EGBd theory allows for traversable wormholes without 
the need for exotic matter, since the quadratic gravitational
terms contribute to the effective stress-energy tensor,
enabling a violation of the energy conditions
(\cite{Kanti:2011jz}).
The domain of existence of the resulting static wormholes
is exhibited in Fig.\,\ref{fig7} (left), where the scaled area
of the throat is shown versus the scaled scalar charge.
At the left boundary wormholes with a double throat
arise, at the lower boundary a transition to black holes
is encountered, while at the right boundary
solutions with a singularity arise.
The lower part of this domain of existence is likely to contain
(mode) stable wormholes. 

EGBd wormholes possess bound states of test particles,
as illustrated in Fig.\,\ref{fig7} (right).
These may reside in either asymptotically flat universe,
but also move between universes.
To obtain a low surface gravity at the throat,
which is on the order of the gravitational
acceleration on the surface of the earth, these wormholes
need an astronomical size with a throat radius on the order of
10 - 100 lightyears. 
Still, smaller EGBd wormholes with higher surface gravity
may be considered as potential astrophysical objects,
whose observational signatures can be studied.

\section{Conclusions and outlook}

We have studied the physical properties of black holes in EGBd theory,
an astrophysically viable alternate theory of gravity.
The coupling constant of the theory has a theoretical bound,
and can be constrained by current and future astrophysical 
observations.

The EGBd black holes differ from their GR counterparts
in a number of significant ways. For instance, 
they possess a minimal size and mass for a given coupling, 
their angular momentum can exceed the Kerr bound, 
and their quadrupole moment can be far bigger.
Also, depending on the coupling to ordinary matter,
the radii and frequencies of their ISCOs can change considerably. 
A general analysis of their geodesics is still missing,
but their shadow is currently being analyzed.
Here further surprises may arise.

We have also analyzed the quasinormal modes of the static
EGBd black holes, leaving the analysis of the rotating black holes
as a challenge for the future.
This analysis leads to the conclusion, that the static EGBd
black holes should be mode stable, since all 
frequencies (calculated) have a negative imaginary part
and thus decay exponentially in time. This is in accordance to the results of an early study
\cite{Kanti2} where a semi-analytic method demonstrated the stability of
these black holes under linear perturbation.
On the other hand, the quasinormal modes with the smallest
imaginary part control the late-time dynamics of perturbed
black holes.

The constraints on the GB coupling are expected to be improved
by future observations
with third generation gravitational wave detectors. 
A simple estimate performed in \cite{Blazquez-Salcedo:2016enn}
leads to the following upper bound on the GB coupling constant
\begin{equation}
\sqrt{\alpha}\lesssim   11 
\left(\frac{50}{\rho}\right)^{1/4}\left(\frac{M}{10 M_\odot}\right)\,{\rm km}
\ ,
\end{equation}
where $\rho$ is the signal-to-noise ratio in the ringdown waveform.
The Einstein telescope may achieve a
signal-to-noise ratio $\rho \approx 100$ for an event like
GW150914.
This would then translate into the bound
\begin{equation}
\sqrt{\alpha}\lesssim   8 \left( \frac{M}{10 M_\odot} \right) \,{\rm km}
\end{equation}
and $\zeta \lesssim 0.4$.

\end{document}